\documentclass[12pt,a4paper]{article}

\def\lvec#1{\setbox0=\hbox{$#1$}
    \setbox1=\hbox{$\scriptstyle\leftarrow$}
    #1\kern-\wd0\smash{
    \raise\ht0\hbox{$\raise1pt\hbox{$\scriptstyle\leftarrow$}$}}
    \kern-\wd1\kern\wd0}
\def\rvec#1{\setbox0=\hbox{$#1$}
    \setbox1=\hbox{$\scriptstyle\rightarrow$}
    #1\kern-\wd0\smash{
    \raise\ht0\hbox{$\raise1pt\hbox{$\scriptstyle\rightarrow$}$}}
    \kern-\wd1\kern\wd0}

\def\diracstar#1#2{
    \setbox0=\hbox{$\gamma$}\setbox1=\hbox{$\gamma_{#1}$}
    \gamma_{#1}\kern-\wd1\kern\wd0
    \smash{\raise4.5pt\hbox{$\scriptstyle#2$}}}

\newcommand{\beq}{\begin{equation}}
\newcommand{\eeq}{\end{equation}}
\newcommand{\beqn}{\begin{eqnarray}}
\newcommand{\eeqn}{\end{eqnarray}}
\newcommand{\nn}{\nonumber}

\date{}
\usepackage{chir_layout}
\begin{document}


\title{\bf On the Geometry of Surface Stress}
\author{
{\large{G.C.~Rossi}}$^{^{a)*)}}$
{\large{M.~Testa}}$^{^{b)**)}}$\\\\
\small $^{a)} $Dipartimento di Fisica, 
Universit\`a di Roma {\it Tor Vergata}\\
\small and INFN, Sezione di Roma 2 \\
\small Via della Ricerca Scientifica, 00133 Roma, Italy\\
\small ${^{b)}} $Dipartimento di Fisica, 
Universit\`a di Roma {\it La Sapienza}\\
\small and INFN, Sezione di Roma {\it La Sapienza}\\
\small P.le A. Moro 2, I-00185 Roma, Italy} 

\maketitle

\begin{abstract}
We present a fully general derivation of the Laplace--Young formula and discuss the interplay between the intrinsic surface geometry and the extrinsic one ensuing from the immersion of the surface in the ordinary euclidean three-dimensional space. We prove that the (reversible) work done in a general surface deformation can be expressed in terms of the surface stress tensor and the variation of the intrinsic surface metric. 
\end{abstract}
\vspace{3.cm}
\noindent\small {$^{*)}${\it Correspondence to:} G.C. Rossi, Phone:
+39-0672594571; FAX: +39-062025259; e-mail address: rossig@roma2.infn.it}\\
\noindent\small {$^{**)}$e-mail address: massimo.testa@roma1.infn.it}
\vfill

\section{Introduction}
\label{sec:INTRO}

The notion of surface tension in fluids dates back to the seminal writings of Laplace~\cite{L} and Young~\cite{Y} where the famous formula relating the difference of the external and internal hydrostatic pressure of a spherical surface to the product of the mean curvature times the surface tension was first derived (see below eq.~(\ref{EQPS1})).  

The concept of surface stress (or interface stress) is of special importance for applications in the interdisciplinary areas of material science, physical chemistry and continuum mechanics~\cite{S1}-\cite{S5} and it has been the subject of extensive studies since when it was introduced by Gibbs~\cite{GIBBS}. A far for complete list of papers and books describing recent investigations in the field can be found in refs.\ from~\cite{REF1} to~\cite{WEISS1}. 

Despite this long history there seem to be still debatable issues and open questions on the subject, like the validity of the Shuttleworth~\cite{REF1} equation or the quest for an  expression of the surface stress in terms of the microscopic degrees of freedom of the system (for instance, of the kind one can write down for the bulk stress tensor, see~\cite{IK,EW1,MIS1,MRT} and references therein). 

In this paper we discuss some geometrical aspects of the notion of surface stress tensor, $\gamma^{\alpha\beta}$ ($\alpha,\beta=1,2$), associated to an arbitrarily shaped interfacial surface. Using methods borrowed from Riemannian geometry, that represents the natural tool to deal with a curved two-dimensional manifold embedded in a three-dimensional (flat) ambient, we derive in full generality a formula that in the isotropic and homogeneous case reduces to the Laplace--Young relation. 

In order to make contact with Thermodynamics we give the expression of the (reversible) work done in the deformation of a generic two-dimensional interface, in terms of the surface stress tensor. The result is similar to the celebrated Shuttleworth~\cite{REF1} formula with the difference arising when deformations orthogonal to the surface are allowed (see eq.~(\ref{NEWMETRIC})).

The outline of the paper is as follows. In order to make the paper self-contained we start in sect.~\ref{sec:GEN} by providing an introduction to the geometrical concepts needed for the present discussion. In sect.~\ref{sec:LYF} we derive the generalized Laplace--Young formula that reduces to the {\it classical} one in the case of isotropic and homogeneous systems. Contact with Thermodynamics is made in sect.~\ref{sec:TC} where we provide the relation between the variation of the (Helmholtz) free energy per unit area under a surface deformation and the surface stress tensor. A few concluding remarks can be found in sect.~\ref{sec:CONC}. In an Appendix for completeness we give a derivation of the Stokes theorem in intrinsic coordinates.

\section{Generalities}
\label{sec:GEN}

Let $\cal S$ be a two-dimensional surface embedded in a Euclidean three-dimensional ambient space, described by the parametric equations
\beq
\vec x  \equiv x_k  (u^\alpha) \vec e_k
\, ,\quad k=1,2,3 \, , \quad \alpha =1,2 
\label{PAREQS}
\eeq
with $(u^1,u^2)$ ranging in a simply connected set, $U[{\cal S}]$, and $\vec e_k$ denoting the  orthonormal vectors of a cartesian reference frame. The two independent vectors 
\beq
\vec x_\alpha(u) \equiv \frac{\partial \vec x(u)}{\partial u^\alpha}\, ,\quad \alpha=1,2 \label{NOTA1}
\eeq
span the tangent plane to $\cal S$ at the point $\vec x=\vec x\,(u^1, u^2)$. The unit normal to $\cal S$ is
\beq
\vec n_{\cal S}=\frac{\vec x_1\wedge \vec x_2}{|\vec x_1\wedge \vec x_2|}\, . \label{NORMS}
\eeq
Given a vector $\vec v$ tangent to $\cal S$, eqs.~(\ref{PAREQS}) and~(\ref{NOTA1}) provide a correspondence between its Riemannian contravariant components, $v^\alpha$, in the curvilinear coordinate system $(u^1,u^2)$, and its cartesian components, $v_k$, in the ambient euclidean space, that reads 
\beq
\vec v = v_k \vec e_k = v^\alpha \vec x_\alpha \, . \label{corresp}
\eeq
The embedding of the surface $\cal S$ defined by eq.~(\ref{PAREQS}) in the euclidean three-dimensional space induces on $\cal S$ the Riemannian metric, $g_{\alpha\beta}$, given by
\beq
g_{\alpha\beta} = \vec x_\alpha\cdot\vec x_\beta \, .
\label{METR}
\eeq
Use of this metric allows to express the scalar product of two tangent vectors to $\cal S$ in terms of their Riemannian contravariant components in the intrinsic form
\beq
\vec v \cdot \vec w =v_i w_i = g_{\alpha\beta} v^\alpha w^\beta \, .
\eeq

\subsection{Principal Curvature}
\label{sec:PC} 

Let ${\cal C}$ be a curve parametrized by $\vec x (\ell)$ with $\ell$ its arclength.
The tangent vector $\vec t (\ell) = {{d\vec x (\ell)} / {d \ell}}$ has unit length, so that its derivative is orthogonal to $\vec t$. We have therefore
\beq
{{d \vec t} \over {d \ell}} = K \vec n \, ,\label{frenet}
\eeq
where the unit vector $\vec n$, orthogonal to $\vec t$, is the so-called principal normal. The proportionality factor, $K\equiv {{1}/{R}}$, defines the curvature at any given point along the curve, with $R$ the curvature radius. If the curve lies in a plane, its principal normal also lies on it.

With reference to the surface $\cal S$ parametrized by eqs.~(\ref{PAREQS}), we remark  that  any plane $\Pi$ containing the normal $\vec n_{\cal S}$ (see eq.~(\ref{NORMS})) identifies a plane curve, ${\cal C}_{\Pi}$, on $\cal S$ called a normal section.  If ${\cal C}_{\Pi}$ is a plane curve, we have 
\beq
\vec t (\ell) = {{d\vec x (\ell)} \over {d \ell}}=\vec x_\alpha(u) \frac{du^\alpha(\ell)}{d\ell} \equiv \vec x_\alpha(u) \dot u^\alpha(\ell)\label{tang}
\eeq
and eq.~(\ref{frenet}) becomes
\beq
{{d \vec t} \over {d \ell}} = K \vec n_{\cal S}\, ,
\eeq
because the normal to ${\cal C}_{\Pi}$ is just $\vec n_{\cal S}$, yielding 
\beq
K = \vec n_{\cal S} \cdot {{d \vec t} \over {d \ell}} \, . \label{TGD1}
\eeq
It is interesting to explicitly compute the derivative of $\vec t(\ell)$ with respect to the arclength parameter. One gets
\beq
\frac{d \vec t(\ell)}{d\ell}=\frac{d }{d\ell} \Big{(}\vec x_\alpha(u)\dot{u}^\alpha(\ell)\Big{)}
= \frac{\partial^2\vec x (u)}{\partial u^\alpha \partial u^\beta}\dot{u}^\alpha(\ell)\dot{u}^\beta(\ell)+
\vec x_\alpha(u)\ddot{u}^\alpha(\ell)\, .\label{TGD}
\eeq
Plugging eq.~(\ref{TGD}) into eq.~(\ref{TGD1}) and taking into account the orthogonality of $\vec x_\alpha$ (and hence of $\vec t(\ell)$) to $\vec n_{\cal S}(\ell)$, one obtains  
\beqn
&&K(\ell)=\vec n_{\cal S}(\ell)\cdot\Big{(} \frac{\partial^2\vec x (u)}{\partial u^\alpha \partial u^\beta}\dot{u}^\alpha(\ell)\dot{u}^\beta(\ell)+
\vec x_\alpha(u)\ddot{u}^\alpha(\ell)\Big{)} = \nn\\
&&\phantom{K(\ell)}=\vec n_{\cal S}(\ell)\cdot\frac{\partial^2\vec x (u)}{\partial u^\alpha \partial u^\beta}\dot{u}^\alpha(\ell)\dot{u}^\beta(\ell) \equiv K_{\alpha\beta}(\ell)\dot{u}^\alpha(\ell)\dot{u}^\beta(\ell) \, ,
\label{KAPC}
\eeqn
where we have introduced the definition 
\beq
K_{\alpha\beta}(\ell)=\vec n_{\cal S}(\ell)\cdot\frac{\partial^2\vec x (u)}{\partial u^\alpha \partial u^\beta}\, .\label{DEFK}
\eeq
$K(\ell)$ is the curvature of the normal section ${\cal C}_{\Pi}$ at the point $u^\alpha=u^\alpha(\ell)$, where the tangent vector has components $\dot{u}^\alpha(\ell)$, $\alpha=1,2$.

$K_{\alpha\beta}$ is a rank two tensor under surface coordinates transformations as it follows by a direct computation. Indeed, taking the derivative of the identity
\begin{equation}
\vec n_{\cal S} (u) \cdot \vec x_\alpha = 0 \, ,\label{IDENTI}
\end{equation}
with respect to $u^\beta$, we have
\begin{equation}
(\partial_{\beta} \vec n_{\cal S}) \cdot \vec x_\alpha + \vec n_{\cal S} (u) \cdot \frac{\partial^2\vec x (u)}{\partial u^\alpha \partial u^\beta} = 0 \, ,
\end{equation}
implying the result 
\beq
K_{\alpha\beta}(u)= - \vec x_\alpha \cdot \partial_{\beta} \vec n_{\cal S} \, ,
\eeq
which proves the thesis.

Eq.~(\ref{DEFK}) defines a real symmetric tensor, $K_{\alpha\beta}=K_{\beta\alpha}$, to which one can associate the two-dimensional eigenvalue problem
\beq
K_{\alpha\beta} {\tau_{(i)}}^\beta =k_i \tau_{(i)\alpha} \, . \label{eigen}
\eeq
The two eigenvalues $k_1={1}/{R_1}$ and $k_2={1}/{R_2}$ define the principal curvature radii $R_1$ and $R_2$ corresponding to the eigenvectors $\tau_{(i)\alpha}$, $i=1,2$, and are the smallest and the largest curvature radii among all the normal sections, as it follows from the elementary inequalities $k_{(2)} ||\chi||^2 \leq \chi^\alpha K_{\alpha\beta}\chi^\beta \leq k_{(1)} ||\chi||^2$ valid for any vector, $\chi^\alpha$, tangent to $\cal S$.

From eq.~(\ref{eigen}) and the symmetry of $K_{\alpha\beta}$ one gets 
\beq
{\tau_{(1)}}^\alpha K_{\alpha\beta} {\tau_{(2)}}^\beta = k_2 \,{\tau_{(1)}}^\alpha  \tau_{(2)\alpha}= k_1\, {\tau_{(2)}}^\alpha  \tau_{(1)\alpha} \, ,
\eeq
which proves the orthogonality relation ${\tau_{(2)}}^\alpha \, \tau_{(1)\alpha}={\tau_{(2)}}^\alpha g_{\alpha\beta} {\tau_{(1)}}^\beta=0$ in the metric~(\ref{METR}) and therefore also when they are considered as vectors in the three-dimensional ambient space, in accordance with eq.~(\ref{corresp}). The eigenvectors $\tau_{(1)}$ and $\tau_{(2)}$ obey the completeness relation
\beq
{\tau_{(1)}}^\alpha\, {\tau_{(1)}}^\beta + {\tau_{(2)}}^\alpha \,{\tau_{(2)}}^\beta = g^{\alpha \beta} \, , \label{complete}
\eeq
as can be checked by taking the scalar product of eq.~(\ref{complete}) with $\tau_{(1)}$ and $\tau_{(2)}$. From eqs.~(\ref{eigen}) and~(\ref{complete}) one gets the well known geometrical result 
\beq
{\rm Tr}[K]\equiv g^{\alpha\beta} K_{\alpha\beta} = {{1} \over {R_1}} + {{1} \over {R_2}} \, .
\eeq

\section{The Laplace--Young formula}
\label{sec:LYF}

\subsection{The general case}
\label{sec:GENCAS}

The description of surface forces requires introducing the two-dimensional (surface) stress tensor $\gamma^{\alpha\beta}$ in analogy with what is done in the three-dimensional bulk when the stress tensor, $\tau_{ik}$, $i,k=1,2,3$, is introduced to describe volume forces~\cite{LL}. 

On the surface ${\cal S}_{\cal C}$ separating two media, the force exerted on the interior of the curve ${\cal C}$ limiting ${\cal S}_{\cal C}$, is given by the formula 
\beq
\vec F({\cal C})=\oint_{{\cal C}} \vec x_\alpha \gamma^{\alpha\beta} n_\beta \,d\ell \, ,
\label{FC}
\eeq
where $n_\beta$ are the components of the unit vector orthogonal to ${\cal C}$, tangent to ${\cal S}_{\cal C}$ and directed towards the interior of ${\cal C}$. Eq.~(\ref{FC}) should be regarded as the definition of the surface stress tensor $\gamma^{\alpha\beta}$. 

This tensor expresses the $\alpha$ component of the force per unit length exerted on an infinitesimal line element whose normal (lying on the tangent plane to the surface) is $n_{\beta}$.

Using the Stokes theorem in intrinsic coordinates~\cite{WALD} (that for completeness we prove in the Appendix), one can rewrite eq.~(\ref{FC}) as a flux integral over a surface, ${\cal S}_{\cal C}$, bounded by ${\cal C}$, in the form
\beq
\vec F({\cal C})=\oint_{{\cal C}} \vec x_\alpha \gamma^{\alpha\beta} n_\beta d\ell = \int_{U({{\cal S}_{\cal C}})}\partial_\beta(\vec x_\alpha \gamma^{\alpha\beta})\,d\sigma\, .
\label{FCF}
\eeq
The equilibrium condition at the interface of two media takes then the expression 
\beq
\int_{U({{\cal S}_{\cal C}})}(\tau^{(2)}_{ik}-\tau^{(1)}_{ik})\,n_{\cal S}^k \, d\sigma = \int_{U({{\cal S}_{\cal C}})}\partial_\beta(\vec x_\alpha \gamma^{\alpha\beta})\,d\sigma 
\, ,\label{EQUINEW}
\eeq
where $\tau^{(2)}_{ik}$ and $\tau^{(1)}_{ik}$ are the bulk stress tensors computed on the two sides of the separating surface. Eq.~(\ref{EQUINEW}) together with~(\ref{FCF}) leads to the local relation
\beq
(\tau^{(2)}_{ik}-\tau^{(1)}_{ik})\,n_{\cal S}^k = \partial^\beta(x_{i}^{\alpha} \gamma_{\alpha\beta})\, ,\label{EQUINE}
\eeq
in agreement with the result derived in ref.~\cite{RUS}. 

Eq.~(\ref{EQUINE}) can be further elaborated by explicitly performing the derivative indicated in its r.h.s. One finds in this way 
\beqn
&&(\tau^{(2)}_{ik}-\tau^{(1)}_{ik})\,n_{\cal S}^k = \partial^\beta(x_{i}^{\alpha} \gamma_{\alpha\beta})=\nn\\&&=  \frac{\partial^2 x_i}{\partial u^\alpha \partial u^\beta}\gamma_{\alpha\beta}+x_i^\alpha\partial^\beta\gamma_{\alpha\beta}=n_{\cal S}^i K^{\alpha\beta} \gamma_{\alpha\beta}+x_i^{\alpha} \nabla^\beta \gamma_{\alpha\beta}\, ,\label{EQUINEWW}
\eeqn
where we have used the fact that, according to eq.~(\ref{DEFK}), $K^{\alpha\beta}$ is the component of the tensor ${\partial^2 \vec x}/{\partial u^\alpha \partial u^\beta}$ along $\vec n_{\cal S}$ and we have introduced the covariant divergence of the surface stress tensor
\beq
\nabla_\beta \gamma^{\alpha\beta} =\partial_\beta \gamma^{\alpha\beta} + \Gamma^\alpha_{\beta\delta} \gamma^{\delta\beta}+ \Gamma^\beta_{\beta\delta}\gamma^{\alpha\delta}
\label{CHR1A}
\eeq
in terms of Christoffel symbols~\cite{LC}.

Projecting eq.~(\ref{EQUINEWW}) along the normal $\vec n_{\cal S}$ and on the plane orthogonal to it, we get the two relations (remember eq.~(\ref{IDENTI}))
\beqn
&&n_{\cal S}^i(\tau^{(2)}_{ik}-\tau^{(1)}_{ik})\,n_{\cal S}^k = \gamma_{\alpha\beta} K^{\alpha\beta}\, ,\label{CLYP1}\\
&&\nabla^\beta \gamma_{\alpha\beta}=0\, ,\label{CLYP2}
\eeqn
that represent two of the key results of this paper. The first equation is the generalization of the equilibrium condition at the interface in the non homogeneous and isotropic case, mi.e.\ the generalized Laplace--Young equation. The second says that the tensor $\gamma_{\alpha\beta}$ is covariantly constant on the surface ${\cal S}$.

\subsection{The isotropic and homogeneous case}
\label{sec:ISOHOM}

The {\it classical} Laplace--Young formula~\cite{L,Y} directly follows from eq.~(\ref{FCF}) in the case of isotropy and homogeneity. In this situation the surface stress tensor has the form $\gamma^{\alpha\beta}= \gamma \, g^{\alpha\beta}$, so the force acting on the surface element $d\sigma$ becomes 
\beq
dF^i= - \Big{[}n_{\cal S}^i {\mbox{Tr}}[K] \gamma +x^i_{\alpha} \partial^\alpha \gamma\Big{]}\,d\sigma\, ,
\label{INF}
\eeq
where we used the relation
\beq
\nabla_\beta \gamma^{\alpha\beta}= \nabla_\beta [\gamma \,g^{\alpha\beta}] = \gamma \nabla_\beta g^{\alpha\beta} + g^{\alpha\beta} \partial_\beta \gamma = \partial^\alpha \gamma \, ,
\label{RELA}
\eeq
that follows from $\nabla_\beta g^{\alpha\beta} =0$. 

The surface element will be in equilibrium if the force $dF^i$ is compensated by the force due to the (normal) pressure difference, $\Delta p = p^{(2)}-p^{(1)}$ of the two media at the interface, i.e.\ if 
\beq
- \vec n_{\cal S} \Delta p+ \vec n_{\cal S} {\mbox{Tr}}[K] \gamma +\vec x_{\alpha} \partial^\alpha \gamma=0\, .
\label{EQP}
\eeq
Projecting out the component of this relation along the normal $\vec n_{\cal S}$ and on the plane orthogonal to it, we get the two scalar relations
\beqn
&&\Delta p= {\mbox{Tr}}[K] \gamma = \gamma \left ( {{1} \over {R_1}} + {{1} \over {R_2}} \right ) \, ,\label{EQPS1}\\
&&\partial_\alpha \gamma=0\, .\label{EQPS2}
\eeqn
The first is the {\it classical} Laplace--Young formula, as first formalized in ref.~\cite{GIB2}, and the second is the known result that says that the surface tension is a constant on the surface ${\cal{S}}$. 

Naturally eqs.~(\ref{EQPS1}) and~(\ref{EQPS2}) are nothing but eqs.~(\ref{CLYP1}) and~(\ref{CLYP2}) in the isotropic and homogeneous limit.

\section{Thermodynamic of a deformation}
\label{sec:TC}

In this section we will consider the work done in a deformation of the equilibrium surface, a notion that is of paramount importance for every thermodynamic application. We start with a brief geometrical introduction.

\subsection{Some geometrical considerations}
\label{sec:GC}

An infinitesimal deformation of ${\cal S}$ can be described by a first order infinitesimal vector, $\delta \vec x(u)$, which gives rise to the displaced surface, ${\cal S}'$ described by the deformed parametric equations
\begin{equation}
\vec x\,'(u) = \vec x(u) + \delta \vec x(u) \label{DEFORM}\, .
\end{equation}
The infinitesimal vector $\delta \vec x(u)$ can be split in the form
\begin{equation}
\delta \vec x(u) = \epsilon (u) \vec n_{\cal S} + \eta^\alpha(u) \vec x_\alpha \, .  \label{DEFORMED}
\end{equation}
We are interested in computing the metric, $g'_{\alpha \beta}$, of the displaced surface, ${\cal S}'$. One finds from the definition~(\ref{METR}) 
\begin{eqnarray}
g'_{\alpha \beta} \approx g_{\alpha \beta} + \vec x_\alpha \cdot \frac {\partial \delta \vec x} {\partial u^\beta} +\vec x_\beta \cdot \frac {\partial \delta \vec x} {\partial u^\alpha} \, .
\end{eqnarray}
Since from eq.~(\ref{DEFORMED}) one finds 
\begin{eqnarray}
\frac {\partial \delta \vec  x} {\partial u^\beta} = 
(\partial_\beta\epsilon (u) )\vec n_{\cal S} + \epsilon (u) \partial_\beta \vec n_{\cal S} + (\partial_\beta \eta^\gamma(u)  ) \vec x_\gamma + \eta^\gamma(u) \partial_\beta \vec x_\gamma
\end{eqnarray}
and
\begin{equation}
\vec x_\alpha \cdot \frac {\partial \delta \vec x} {\partial u^\beta} = \epsilon (u)  \vec x_\alpha \cdot \partial_\beta \vec n_{\cal S} +
(\partial_\beta \eta^\gamma(u)) \vec x_\alpha \cdot \vec x_\gamma + \eta^\gamma(u) \vec x_\alpha \cdot \partial_\beta \vec x_\gamma \, ,
\end{equation}
one gets (see eq.~(\ref{PROJ2}))
\begin{eqnarray}
&&\vec x_\alpha \cdot \frac {\partial \delta \vec x} {\partial u^\beta} = - \epsilon (u) K_{\alpha\beta}(u)
+  g_{\alpha \gamma} \partial_\beta \eta^\gamma(u) + \eta^\gamma(u) {\Gamma^\delta}_{\beta \gamma} g_{\delta \alpha} \equiv \nonumber\\
&&\phantom{\vec x_\alpha \cdot \frac {\partial \delta \vec x} {\partial u^\beta}}\equiv  - \epsilon (u) K_{\alpha\beta}(u)
+  g_{\alpha \gamma} \partial_\beta \eta^\gamma(u) + \eta^\gamma(u) \left[ \beta \gamma, \alpha  \right ] =\nonumber \\
&&\phantom{\vec x_\alpha \cdot \frac {\partial \delta \vec x} {\partial u^\beta}}=  - \epsilon (u) K_{\alpha\beta}(u) + \nabla_\beta \eta_\alpha \, ,
\end{eqnarray}
which finally yields  
\begin{equation}
\delta g_{\alpha \beta}= g'_{\alpha \beta} -g_{\alpha \beta} \approx - 2 \epsilon (u) K_{\alpha\beta}(u) + \nabla_\beta \eta_\alpha+ \nabla_\alpha \eta_\beta \, . \label{NEWMETRIC}
\end{equation}

\subsection{Work and free energy}
\label{sec:WFE}

We are now ready to compute the work, $\delta W$, performed by the surface stress under the infinitesimal deformation~(\ref{DEFORMED}). Recalling that $\delta W$ has two contributions, one from the stretching of the boundary curve, $\partial {\cal S}$, and another one from the ``bulk'' deformation of ${\cal S}$ itself, one gets using eq.~(\ref{DIVS})
\begin{eqnarray}
\delta W = - \int_{\cal S} \left [ \epsilon \, K_{\alpha\beta} \gamma^{\alpha\beta} + \eta_\alpha \nabla_\beta \gamma^{\alpha\beta} \right ] d \sigma - \int_{\partial {\cal S}} \gamma_{\alpha \beta}\, \eta^\alpha n^\beta d \ell \, ,\label{WORK0}
\end{eqnarray}
From the Stokes theorem in intrinsic coordinates~\cite{WALD} (see Appendix), we obtain 
\begin{equation}
 \int_{\partial {\cal S}} \gamma_{\alpha \beta} \eta^\alpha n^\beta d \ell = - \int_{\cal S} \nabla^\alpha (\gamma_{\alpha \beta} \eta^\beta) d \sigma \, , \label{GAUSS}
\end{equation}
with the minus sign due to the orientation of the surface normal $n$, that in our convention is directed towards the interior of $\partial {\cal S}$. In virtue of eq.~(\ref{GAUSS}), eq.~(\ref{WORK0}) becomes
\begin{eqnarray}
\delta W = - \int_{\cal S} \left [ \epsilon \, K_{\alpha\beta} \gamma^{\alpha\beta} + \eta_\alpha \nabla_\beta \gamma^{\alpha\beta} \right ] d \sigma + \int_{{\cal S}} \nabla^\beta (\gamma_{\alpha \beta} \eta^\alpha) d \sigma = \nonumber\\ = \int_{\cal S} \left [ - \epsilon \, K_{\alpha\beta} \gamma^{\alpha\beta} + \gamma^{\alpha\beta} \nabla_\beta \eta_\alpha \right ] d \sigma = {{1} \over {2}} \int_{\cal S} \delta g_{\alpha\beta} \gamma^{\alpha\beta} d \sigma \, ,\label{WORK2}
\end{eqnarray}
where $\delta g_{\alpha\beta}$ is the variation of the surface metric (eq.~(\ref{NEWMETRIC})) under the deformation~(\ref{DEFORMED}). The final formula
\begin{eqnarray}
\delta W ={{1} \over {2}} \int_{\cal S} \delta g_{\alpha\beta} \gamma^{\alpha\beta} d \sigma \, ,\label{WORK3}
\end{eqnarray}
is very interesting because it allows us to derive a thermodynamic definition of surface stress. In fact, under the assumption that the surface deformation~(\ref{DEFORM}) is carried out reversibly, one can identify $\delta W$ with minus the (Helmholtz) free energy variation, $-\delta A$. Recalling eqs.~(\ref{COMB1}) and~(\ref{DETG}), one can derive from eq.~(\ref{WORK2}) the local equation  
\beq
-\frac{\delta A}{\delta g_{\alpha\beta}}=\frac{1}{2}\sqrt{|\det g|}\,\gamma^{\alpha\beta}\, .
\label{SURSTR}
\eeq
If, as it is customary, one introduces the free energy per unit area, $a\equiv A /\sigma_{\cal S}$, from eq.~(\ref{SURSTR}) one obtains 
\beq
\gamma^{\alpha\beta}=-\frac{2}{\sqrt{|\det g|}}\frac{\delta (a\sigma_{\cal S})}{\delta g_{\alpha\beta}}=-a \,g^{\alpha\beta}- 2\frac{\sigma_{\cal S}}{\sqrt{|\det g|}}\frac{\delta a}{\delta g_{\alpha\beta}}\, .
\label{SURSTRU}
\eeq
This equation is reminiscent of the Shuttleworth formula~\cite{REF1}, but not identical with it. A part from the trivial fact that eq.~(\ref{SURSTRU}) correctly takes into account the general tensor nature of the surface stress~\footnote{In the case of an isotropic medium, $\gamma^{\alpha\beta}= \gamma \, g^{\alpha\beta}$, eq.~(\ref{WORK2}) can be written in the form 
\begin{eqnarray}
\delta W_{isotropic} = \int_{{\cal S}'} \gamma  d \sigma - \int_{{\cal S}} \gamma  d \sigma \nn
\end{eqnarray}
which, in the case of a constant surface stress, $\gamma$, across ${\cal S}$, becomes
\begin{eqnarray}
\delta W_{isotropic} =  \gamma \left [\int_{{\cal S}'} d \sigma - \int_{{\cal S}} d \sigma \right ] =\gamma \, \delta\sigma_{\cal S} \, ,\nn
\end{eqnarray}
$\delta \sigma_{\cal S}$ being the variation of the area of $\cal S$, in agreement with the usual definition of isotropic surface stress.}, the crucial difference is that the derivative of the free energy per unit area is taken in eq.~(\ref{SURSTRU}) with respect to the metric tensor $g_{\alpha\beta}$ and not with respect to the strain tensor $(\nabla_{\alpha}\eta_{\beta}+\nabla_{\beta}\eta_{\alpha})/2$, as it is done in ref.~\cite{REF1} and in all the subsequent literature. As it is clear from eq.~(\ref{NEWMETRIC}), (variations under the) metric tensor and strain tensor do not coincide, unless $\epsilon(u)=0$. 

As a side remark, we note that the idea of defining the bulk (three-dimensional) stress tensor as the response of the free energy under a deformation of the (three-dimensional) metric was advocated in refs.~\cite{MIS1,MIS2}. In that case, however, it was shown~\cite{MRT,RT,MRTC} that derivatives with respect to the (three-dimensional) deformation tensor and derivatives with respect to the (three-dimensional) ambient space metrics give identical results.

\section{Conclusions}
\label{sec:CONC}

Using elements of Riemannian tensor calculus, we have given a geometrical characterization of the surface stress tensor and rederived the Laplace--Young formula for an arbitrarily curved interfacial surface. We have also discussed the expression of the (reversible) work done in a surface deformation and we have shown that it is given by a two-dimensional integral where the surface stress tensor is saturated with the deformation of the surface intrinsic metric tensor (and not with the strain tensor).

The computation of $\gamma^{\alpha\beta}$ in terms of the statistical mechanics of the microscopic degrees of freedom of the system is still an open problem and will be the object of further studies.

\appendix 
\section*{Appendix - The Stokes theorem in intrinsic coordinates}  
\renewcommand{\thesection}{A} 
\label{sec:APPA}

For completeness we give in this Appendix a ``poor-man'' derivation of the Stokes theorem in intrinsic coordinates~\cite{WALD}. More precisely for the needs of this paper we want to prove the formula
\beq
\oint_{{\cal C}} x^i_\alpha \chi^{\alpha\beta} n_\beta \,d\ell=- \int_{U[{\cal{S_{\cal C}}}]}\partial_\beta(x^i_\alpha\chi^{\alpha\beta}) \,d\sigma\, ,\quad 1=1,2,3\, ,
\label{DIVSL}
\eeq
where $\chi^{\alpha\beta}$ is a two-dimensional tensor and ${\cal{S_{\cal C}}}$ is an open surface, bounded by the smooth curve ${\cal C}$, with parametric coordinates $(u^1,u^2)$ spanning the two-dimensional set $U[{\cal{S_{\cal C}}}]$. 

The proof of~(\ref{DIVSL}) relies on the use of the standard Stokes theorem (in extrinsic coordinates)
\beq
\oint_{{\cal C}} \vec v \cdot \vec t \,d\ell= \int_{{\cal S}_{\cal C}} {\mbox{rot}}\,\vec v \cdot \vec n_{\cal S}\,d\sigma\, ,
\label{STOK}
\eeq
with $\vec v$ a smooth vector field and $\vec n_{\cal S}$ the oriented normal to the surface ${\cal S}_{\cal C}$. Signs are chosen so that the curve ${\cal C}$ winds counter-clockwise with respect to $\vec n_{\cal S}$. 

The first step of our proof consists in rewriting the surface integral in the r.h.s.\ of eq.~(\ref{STOK}) in the form 
\begin{equation}
\int_{S_{\cal C}} {\mbox{rot}}\,\vec v \cdot \vec n_{\cal S}d\sigma=\int_{U[{\cal{S_{\cal C}}}]} \Big{(} \frac{\partial (\vec v\cdot \vec x_2)}{\partial u^1}-\frac{\partial (\vec v\cdot \vec x_1)}{\partial u^2}\Big{)}\, du^1du^2 \, ,\label{PASS}
\end{equation}
where use has been made of the elementary formulae
\beqn
&&\vec n_{\cal S}d\sigma=(\vec x_1\wedge \vec x_2)\,du^1du^2\, ,\label{COMB1}\\
&& |\vec x_1\wedge \vec x_2|=\sqrt {|{\mbox{det}}\,g|} \, ,\label{DETG}\\
&&[{\mbox{rot}}\,\vec v]_i=\epsilon_{ijk}\frac{\partial v_k}{\partial x_j} \, ,\label{COMB2}
\eeqn
and we have employed the definition ${\mbox{det}\,g}=\mbox{det}_{\alpha\beta}\,g_{\alpha\beta}$. 

With the identifications for fixed $i=1,2,3$
\beqn
&&\vec v \to x^i_\alpha \chi^{\alpha\beta}(\vec x_\beta \wedge\vec n_{\cal S}) \, ,\\
\label{IDENT}
&&\vec v\cdot \vec x_1 \to \sqrt{|{\mbox{det}}\,g|}\,x^i_\beta\chi^{\beta 2}\, ,\label{VX1}\\
&&\vec v\cdot \vec x_2 \to - \sqrt{|{\mbox{det}}\,g|}\,x^i_\beta\chi^{\beta 1}\, ,\label{VX2}
\eeqn 
we find the correspondences
\beqn
\hspace{-1.7cm}&&\oint_{{\cal C}} \vec v \cdot \vec t \,d\ell \to \oint_{{\cal C}} x^i_\alpha \chi^{\alpha\beta}[(\vec x_\beta \wedge \vec n_{\cal S})\cdot \vec t] d\ell\, ,\label{COR1}\\
\hspace{-1.7cm}&&\int_{U[{\cal{S_{\cal C}}}]} \!\Big{(} \frac{\partial (\vec v\cdot \vec x_2)}{\partial u^1}-\frac{\partial (\vec v\cdot \vec x_1)}{\partial u^2}\Big{)}du^1du^2 \to - \int_{U[{S_{\cal C}}]}\!\! \frac{\partial (\sqrt{|{\mbox{det}}\,g|}\,x^i_\alpha\chi^{\alpha\beta})}{\partial u^\beta}du^1du^2\, ,\label{COR2}
\eeqn
hence, using~(\ref{STOK}) and~(\ref{PASS}), we get the equality
\beqn
\hspace{-1.7cm}&&\oint_{{\cal C}} x^i_\alpha \chi^{\alpha\beta}[(\vec x_\beta \wedge \vec n_{\cal S})\cdot \vec t] d\ell=- \int_{U[{S_{\cal C}}]}\!\! \frac{\partial (\sqrt{|{\mbox{det}}\,g|}\,x^i_\alpha\chi^{\alpha\beta})}{\partial u^\beta}du^1du^2\, ,\label{COR3}
\eeqn
from which eq~(\ref{DIVSL}) follows. In fact, we find for the l.h.s.
\beqn
\hspace{-1.7cm}&&\oint_{{\cal C}} x^i_\alpha \chi^{\alpha\beta}[(\vec x_\beta \wedge \vec n_{\cal S})\cdot \vec t] d\ell=\oint_{{\cal C}} x^i_\alpha \chi^{\alpha\beta} [\vec x_\beta \cdot (\vec n_{\cal S} \wedge\vec t)] d\ell=\oint_{{\cal C}} x^i_\alpha \chi^{\alpha\beta} n_\beta \,d\ell\, ,
\label{COR4}
\eeqn
where in the last equality we have used the relations 
\beqn 
&& \vec n= \vec n_{\cal S} \wedge\vec t \, .\label{NSNT}\\
&&n_\beta= \vec n \cdot \vec x_\beta \, .\label{NDNT}
\eeqn
Explicitly performing the derivative in the r.h.s.\ of eq.~(\ref{COR3}), we obtain
\beqn
\hspace{-.8cm}&&\int_{U[{S_{\cal C}}]} \frac{\partial (\sqrt{|{\mbox{det}}\,g|}\,x^i_\alpha\chi^{\alpha\beta})}{\partial u^\beta}\,du^1du^2= \label{DIV}\\
\hspace{-.8cm}&&=\int_{U[{\cal{S_{\cal C}}}]} \Big{[}\frac{\partial \sqrt{|{\mbox{det}}\,g|}}{\partial u^\beta}x^i_\alpha\chi^{\alpha\beta}+\sqrt{|{\mbox{det}}\,g|}\,\frac{\partial^2 x^i}{\partial u^\alpha \partial u^\beta}\chi^{\alpha\beta}+\sqrt{|{\mbox{det}}\,g|}\,x^i_{\alpha}\frac{\partial\chi^{\alpha\beta}}{\partial u^\beta}\Big{]}\,du^1du^2\, .\nn
\eeqn
In order to bring this formula in the desired form, we observe that the first term in the r.h.s.\ can be rewritten as
\beq
\frac{\partial \sqrt{|{\mbox{det}}\,g|}}{\partial u^\beta}=\Gamma^\delta_{\delta\beta}\sqrt{|{\mbox{det}}\,g|}\, ,
\label{CHR}
\eeq
where $\Gamma^\delta_{\alpha\beta}$ is the Christoffel symbol of the second kind. To analyze  the second term, we introduce the decomposition
\beq
\frac{\partial^2 x^i}{\partial u^\alpha \partial u^\beta}= c^\gamma_{\alpha\beta}x^i_{\gamma}+h_{\alpha\beta} n_{\cal S}^i \, .
\label{DEC}
\eeq
Projecting this relation on the normal $\vec n_{\cal S}$ and on the plane orthogonal to it allows the computation of $c^\gamma_{\alpha\beta}$ and $h_{\alpha\beta}$ for which one finds (recall eq.~(\ref{DEFK}))
\beqn
&& h_{\alpha\beta} =\vec n_{\cal S}\cdot \frac{\partial^2\vec x}{\partial u^\alpha \partial u^\beta}=K_{\alpha\beta} \, ,\label{PROJ1}\\
&&c^\gamma_{\alpha\beta} \vec x_{\gamma} \cdot \vec x_{\delta} = c^\gamma_{\alpha\beta} g_{\gamma \delta} = \vec x_\delta\cdot \frac{\partial^2\vec x}{\partial u^\alpha \partial u^\beta} \equiv {\Gamma^\gamma}_{\alpha \beta}\, g_{\gamma \delta} \, , \label{PROJ2}
\eeqn
where ${\Gamma^\gamma}_{\alpha \beta} \,g_{\gamma\delta}$ denote the Christoffel symbols of the first kind~\cite{LC}. Using eqs.~(\ref{PROJ1}) and~(\ref{PROJ2}), one can rewrite eq.~(\ref{DEC}) as
\beq
\frac{\partial^2 x^i}{\partial u^\alpha \partial u^\beta}= \Gamma^\gamma_{\alpha\beta}x^i_{\gamma}+K_{\alpha\beta} n_{\cal S}^i \, . \label{DEC1}
\eeq
Plugging eqs.~(\ref{CHR}) and~(\ref{DEC1}) into eq.~(\ref{DIV}), we get
\beqn
\hspace{-.4cm}
- \int_{U[{S_{\cal C}}]} \!\!\frac{\partial (\sqrt{|{\mbox{det}}\,g|}\,x^i_\alpha\chi^{\alpha\beta})}{\partial u^\beta}\,du^1du^2= - \int_{U[{\cal{S_{\cal C}}}]} \Big{[}n_{\cal S}^i K_{\alpha\beta} \chi^{\alpha\beta}+x^i_{\alpha} \nabla_\beta \chi^{\alpha\beta}\Big{]}\,d\sigma \, ,\label{DIVS}
\eeqn
where  
\beq 
\nabla_\beta \chi^{\alpha\beta} =\partial_\beta \chi^{\alpha\beta} + \Gamma^\alpha_{\beta\delta} \chi^{\delta\beta}+ \Gamma^\beta_{\beta\delta}\chi^{\alpha\delta}
\label{CHR1}
\eeq 
is the covariant divergence of the $\chi^{\alpha\beta}$ tensor and $d\sigma=\sqrt{|{\mbox{det}}\,g|}\,du^1du^2$ is the (reparametrization invariant) surface element. Noting that the two terms inside the integral~(\ref{DIVS}) can be exactly rearranged into a total divergence, giving
\beq
n_{\cal S}^i K_{\alpha\beta} \chi^{\alpha\beta}+x^i_{\alpha} \nabla_\beta \chi^{\alpha\beta}=\partial_\beta(x^i_\alpha\chi^{\alpha\beta})\, ,
\label{DIVSL1}
\eeq
we simply get 
\beqn
\hspace{-.4cm}
- \int_{U[{S_{\cal C}}]} \!\!\frac{\partial (\sqrt{|{\mbox{det}}\,g|}\,x^i_\alpha\chi^{\alpha\beta})}{\partial u^\beta}\,du^1du^2= - \int_{U[{\cal{S_{\cal C}}}]} \partial_\beta(x^i_\alpha\chi^{\alpha\beta})\,d\sigma \, ,\label{COR5}
\eeqn
Putting together eqs.~(\ref{COR3}), (\ref{COR4}) and~(\ref{COR5}), we obtain announced Stokes theorem in intrinsic coordinates provided by eq.~(\ref{DIVSL}).

\end{document}